\documentclass[onecolumn,floatfix,superscriptaddress,showpacs,showkeys,nofootinbib,preprint]{revtex4}
\usepackage{epsfig}
\usepackage{amssymb,latexsym,amsmath}
\newcommand{\eq}[1]{\begin{align} #1 \end{align}}

\def\d{{\rm d}}

\def\e{{\rm e}}
\def\p{{\rm p}}
\def\i{{\rm i}}

\def\Oav{{\langle O \rangle}}
\def\Nj{{\{N_j\}}}

\def\Omj{\Omega_{\{N_j\}}}
\def\Pij{\Pi_{\{N_j\}}}

\def\Qz{{\bf Q}}

\def\qj{{\bf q}_j}
\def\delq{{\delta_{\Qz,\sum_j N_j \qj}}}

\def\phivs{\mbox{\boldmath${\scriptstyle{\phi}}$}}
\def\pivs{\mbox{\boldmath${\scriptstyle{\pi}}$}}

\begin{document}

\title{THRESHOLD EFFECTS IN RELATIVISTIC GASES}

\author{V.V. Begun}
\affiliation{Bogolyubov Institute for Theoretical Physics, Kiev,
Ukraine}
\author{L. Ferroni}
\affiliation{Universit\`a di Firenze and INFN Sezione di Firenze}
\author{M.I. Gorenstein}
\affiliation{Bogolyubov Institute for Theoretical Physics, Kiev,
Ukraine}
%\affiliation{Institut f\'ur Theoretische Physik,
%Universit\'at Frankfurt, Germany}
\affiliation{Frankfurt Institute
for Advanced Studies, Frankfurt, Germany}
\author{M. Ga\'zdzicki}
\affiliation{Institut f\'ur Kernphysik, Johann Wolfgang
 Goethe Universit\'at Frankfurt, Germany}
\affiliation{\'Swi\c{e}tokrzyska Academy, Kielce, Poland}
\author{F. Becattini}
\affiliation{Universit\`a di Firenze and INFN Sezione di Firenze}

\begin{abstract}
Particle multiplicities and ratios in the microcanonical 
ensemble of relativistic gases near production thresholds are studied. 
It is shown that the ratio of heavy to light particle multiplicity
may be enhanced in comparison to its thermodynamic limit.
\end{abstract}


%\pacs{} \keywords{micro-canonical ensemble, small systems...}
\maketitle

\section{Introduction}

The statistical model of hadron production in high energy collisions of
both elementary and heavy ion collisions proved to be remarkably succesful
in reproducing the multiplicities of different hadron species. This
finding has raised the question of the meaning of thermodynamical quantities
in these systems and triggered a still ongoing debate \cite{becheinz,vari}. 

In the case of relativistic heavy ion collisions, due the large multiplicities
involved, the use of the grand canonical ensemble (GCE) is sufficient and the 
model predicts that hadron yields should be linearly dependent on the volume 
at hadronization. On the other hand, in
elementary collisions, the canonical ensemble (CE) \cite{becheinz,ce-b,ce-c,ce-d,ce-e} 
enforcing the exact conservation of charges must be used owing to the relatively
low multiplicities. At lower multiplicities, even the micro-canonical ensemble 
(MCE), where energy-momentum conservation is enforced, is needed to correctly
describe particle abundances. The transition from MCE to CE has been studied in 
detail in refs.~\cite{shahme1,shahme2}. The equivalence between the three ensembles,
with respect to particle multiplicities, is recovered only in the thermodynamic
limit, when $V \to \infty$. 

The charged conservation laws in the CE imply a reduction of  the mean multiplicity 
of heavy charged hadrons in comparison to the corresponding multiplicity calculated 
in the GCE. This effect is usually defined as {\em canonical suppression} and 
was particularly studied for the production of strange \cite{ce-c} and open charm 
hadrons \cite{ce-e} as well as anti-baryons \cite{ce-d}.

The canonical suppression is significant for the charged particles with a mean 
multiplicity smaller than 1. Naively one may expect that an introduction of the 
energy conservation law by use of the MCE should lead to the gradually increasing
suppression of the multiplicity of heavy hadrons with decreasing energy of the system.
Contrary to this expectation, in this work we demonstrate that the energy and momentum 
conservation laws lead to rich structures in the dependence of hadron multiplicities on
the system energy. 

The paper is organized as follows: in Section II we study this effect by means of 
a simple system of massless and heavy particles, where an analytical description
is possible. In Sect. III, we study the full ideal hadron-resonance gas by means
of numerical methods described in refs.~\cite{shahme1,shahme2}. Conclusions are
summarized in Sect. IV.

\section{The Analytical Model}\label{sec2}

In this section we consider the system of non­interacting
Boltzmann particles with zero mass, both neutral (with degeneracy
factor $g$) and charged (with degeneracy factor $g_{\pm}$). The
system also includes neutral heavy particles with mass $M$ and
degeneracy factor $G$. 
The non­relativistic approximation is used for heavy particles.  

\subsection{Grand Canonical Ensemble}
In the
GCE  independent variables
are volume $V$,  temperature $T$, and chemical potential $\mu$.
The average number of massless neutral particles $N$, massless
charged particles, $N_+$ and $N_-$ , and heavy neutral particles, $n$, 
are:
\eq{ \label{GCE mult}
 \langle N\rangle_{GCE} &\,=\, \frac{gV\,T^3}{\pi^2}~,\quad
 \langle N_{\pm}\rangle_{GCE} \,=\, \frac{g_{\pm}V\,T^3}{\pi^2}\,
 \exp\left(\pm\frac{\mu}{T}\right),\nonumber \\ 
 \langle n\rangle_{GCE}
 &\,=\, GV~\left(\frac{MT}{2\pi}\right)^{3/2}\,
 \exp\left(-\frac{M}{T}\right)~ .
 }
The system energy reads:
\eq{ \label{GCE Energy}
 \langle E\rangle_{GCE} \,\equiv\, \varepsilon(T )\,V
 \,=\, 3T\langle N\rangle_{GCE} + 3T\langle N_+\rangle_{GCE}
     + 3T\langle N_-\rangle_{GCE}
     + \left(\frac{3}{2}T + M\right)\langle n\rangle_{GCE}.
 }
Of course, the
non­relativistic approximation used for massive particles holds
for  $M\gg T$.

\subsection{Micro-Canonical Ensemble}
In %micro-canonical ensemble (
the MCE independent variables are
volume $V$, energy $E$, and net charge $Q$.
The MCE partition function $\Omega_N(E,V)$ in
the system of $N$ neutral massless particles equals to \cite{mce1} :
\eq{ \label{WN(E,V)}
 \Omega_N(E,V) \,=\, \frac{1}{N!}\left(\frac{gV}{2\pi^2}\right)^N
 \int_0^{\infty} p_1^2 dp_1\ldots \int_0^{\infty}p^2_N dp_N\,
 \delta\left(E-\sum^N_{j=1}p_j\right)
 \,=\, \frac{1}{N!}\left(\frac{gV}{\pi^2}\right)^N \frac{E^{3N-1}}{(3N - 1)!}\,.
 }
Let us consider the case of charged massless particles
with total zero net charge, $Q=0$.  This implies
$\mu_Q=0$ in the GCE and thus $\langle
N_+\rangle_{GCE}=\langle N_-\rangle_{GCE}$. The MCE
partition function for $N_+=N_-$ charged massless particles reads
\cite{mce1}:
 \eq{ \label{WNpm(E,V)}
\Omega_{N_{\pm}}(E,\,V)
 \,=\, \frac{1}{N_{\pm}!^2} \left(\frac{g_{\pm}V}{\pi^2}\right)^{2N_{\pm}}
       \frac{E^{6N_{\pm}-1}}{(6N_{\pm}-1)!}\, ,
 }
where $N_{\pm}\equiv N_+=N_-$.

The MCE partition function for $n$ heavy non­relativistic particles
can be  calculated analytically:
 \eq{ \label{Wn(E,V)}
 \Omega_n(E,\,V )
 &\,\equiv\, \frac{1}{n!} \left(\frac{GV}{2\pi^2}\right)
    \int_0^{\infty} p_1^2 dp_1 \ldots \int_0^{\infty}p_n^2 dp_n\,
    \delta\left[E \,-\, \sum_{j=1}^{n}
    \left(M\,+\,\frac{p_j^2}{2M}\right)\right]\nonumber  \\
 &\,=\, \frac{1}{n!}\left(\frac{GV}{(2\pi)^{3/2}}\right)^n
        \frac{M^{\frac{3n}{2}}}{\Gamma\left(\frac{3n}{2}\right)}
        \left(E -nM\right)^{\frac{3n}{2}-1}\,,
 }
where the Euler gamma function $\Gamma(x)$ has a simple form for
integer $k$ and half­integer $k+1/2$ arguments:
\[ \Gamma(k)\,=\,(k - 1)!\,,\quad \Gamma\left(k+\frac{1}{2}\right)
  \,=\, \frac{1\cdot3\cdot \ldots \cdot (2k - 1)}{2^n}\, \sqrt{\pi}\,.
\]
Using Eqs.~(\ref{WN(E,V)}, \ref{Wn(E,V)}) one can calculate the
partition function for $N$ massless and $n$ heavy particles: 
 \eq{ \label{WN,n(E,V)}
 \Omega_{N,n}(E,\,V)
 &\,=\, \int_0^{\infty}dE_1 \int_0^{\infty}dE_2\,
        \Omega_N(E_1,\,V)\,\Omega_n(E_2,V) \delta[E-E_1-E_2] \nonumber \\
 &\,=\, \frac{1}{N!}\, \frac{1}{n!} \left(\frac{gV}{\pi^2}\right)^N
        \left(\frac{GV}{(2\pi)^{3/2}}\right)^n
        \frac{M^{\frac{3n}{2}}}
    {\Gamma(3N+\frac{3}{2}\,n)}\,(E -nM)^{3N+\frac{3n}{2}-1}\,.
        }
\noindent
Note that the MCE partition functions $\Omega_N(E,\,V)$ (\ref{WN(E,V)})
and $\Omega_n(E,V)$ (\ref{Wn(E,V)}) are defined for non­zero particle
numbers, $N\geq 1$ and $n\geq 1$, because of an exact energy
conservation. However, the MCE partition functions
$\Omega_{N,n}(E,\,V)$ (\ref{WN,n(E,V)}) requires only $N+n\geq 1$, so
that either $N$ or $n$ can be equal to zero.

Finally, using Eqs.~(\ref{WNpm(E,V)}, \ref{WN,n(E,V)}) one finds
the partition function for $N$ massless neutral, $N_+=N_-$
massless charged, and $n$ heavy particles:
 \eq{ \label{WN,Npm,n(E,V)}
 \Omega_{N,N_{\pm},n}(  E,\,V)
  &\,=\, \int_0^{\infty} dE_1 \int_0^{\infty}dE_2\,
        \Omega_{N,n}(E_1,\,V)\, \Omega_{N_{\pm}}(E_2,\,V)\,
        \delta[E-E_1-E_2] \nonumber \\
 &\,=\, \frac{1}{N!}\, \frac{1}{(N_{\pm})!^2}\, \frac{1}{n!}
        \left(\frac{gV}{\pi^2}\right)^N
        \left(\frac{g_{\pm}V}{\pi^2}\right)^{2N_{\pm}}
        \left(\frac{GV}{(2\pi)^{3/2}}\right)^n \times \nonumber \\
     &\,   \times \,\frac{M^{\frac{3n}{2}}}{\Gamma(3N+6N_{\pm}+\frac{3}{2}n)}\,
        (E-nM)^{3N+6N_{\pm}+\frac{3}{2}n-1}\, .
 }
The MCE partition function $\Omega_{N,N_{\pm},n}(E,\,V)$
(\ref{WN,Npm,n(E,V)}) is evidently transformed into
$\Omega_{N,n}(E,\,V)$ (\ref{WN,n(E,V)}) for $g_{\pm}=0$. For $g=0$ the
$\Omega_{N,N_{\pm},n}(E,\,V)$ from Eq.~(\ref{WN,Npm,n(E,V)}) is
transformed into the partition function for $N_+=N_-$ massless
charged and $n$ heavy particles:
 \eq{ \label{WNpm,n(E,V)}
 \Omega_{N_{\pm},n}(E,\,V)
 \,=\, \frac{1}{(N_{\pm}!)^2}\,\frac{1}{n!} \left(\frac{g_{\pm}V}{\pi^2}\right)^{2N_{\pm}}
       \left(\frac{GV}{(2\pi)^{3/2}}\right)^n
       \frac{M^{\frac{3n}{2}}}{\Gamma(6N_{\pm}+\frac{3}{2}n)}\,
       (E-nM)^{6N_{\pm}+\frac{3}{2}n-1}\, .
 }

The MCE partition function can also be calculated for the case of
non-zero system net charge $Q$ in two steps:
 \eq{
 \Omega_{N_+,N_-}(E,\,V)
  \,=\, \int_0^{\infty} dE_1 \int_0^{\infty}dE_2\,
        \Omega_{N_+}(E_1,\,V)\,\Omega_{N_-}(E_2,\,V)\,\delta[\,N_+-N_--Q\,]\,
        \delta[\,E-E_1-E_2\,]\,,
 }
 \eq{
 \Omega_{N_+,N_-,n}(E,\,V)
  \,=\, \int_0^{\infty} dE_1 \int_0^{\infty}dE_2\,
        \Omega_{N_+,N_-}(E_1,\,V)\,\Omega_{n}(E_2,\,V)\,\delta[\,E-E_1-E_2\,]\,.
 }
Finally, the partition function for $N_+$ positively charged
massless particles and for $n$ neutral heavy particles equals to:
 \eq{ \label{WN+,n(E,V,Q)}
 \Omega_{N_+,n}(E,\,V,\,Q)
 &\,=\, \frac{1}{N_+!}\,\frac{1}{(N_+-Q)!}\,\frac{1}{n!}
       \left(\frac{g_{\pm}V}{\pi^2}\right)^{2N_+-Q}
       \left(\frac{GV}{(2\pi)^{3/2}}\right)^n \times \nonumber \\
     &\, \times \,\frac{M^{\frac{3n}{2}}}{\Gamma(6N_+-3Q+\frac{3}{2}n)}\,
       (E-nM)^{6N_+-3Q+\frac{3}{2}n-1}\, ,
 }
and for negatively charged massless particles:
 \eq{ \label{WN-,n(E,V,Q)}
 \Omega_{N_-,n}(E,\,V,\,Q)
 &\,=\, \frac{1}{N_-!}\,\frac{1}{(N_-+Q)!}\,\frac{1}{n!}
       \left(\frac{g_{\pm}V}{\pi^2}\right)^{2N_-+Q}
       \left(\frac{GV}{(2\pi)^{3/2}}\right)^n \times \nonumber \\
    &\, \times \,   \frac{M^{\frac{3n}{2}}}{\Gamma(6N_-+3Q+\frac{3}{2}n)}\,
       (E-nM)^{6N_-+3Q+\frac{3}{2}n-1}\, .
 }
These formulas transform into $\Omega_{N_{\pm},n}(E,\,V)$
(\ref{WNpm,n(E,V)}) for
$Q=0$.
%
%******************************************************************
%\section{PARTICLE MULTIPLICITIES IN THE MCE AND GCE}
%******************************************************************

\begin{figure}[h!]
 \epsfig{file=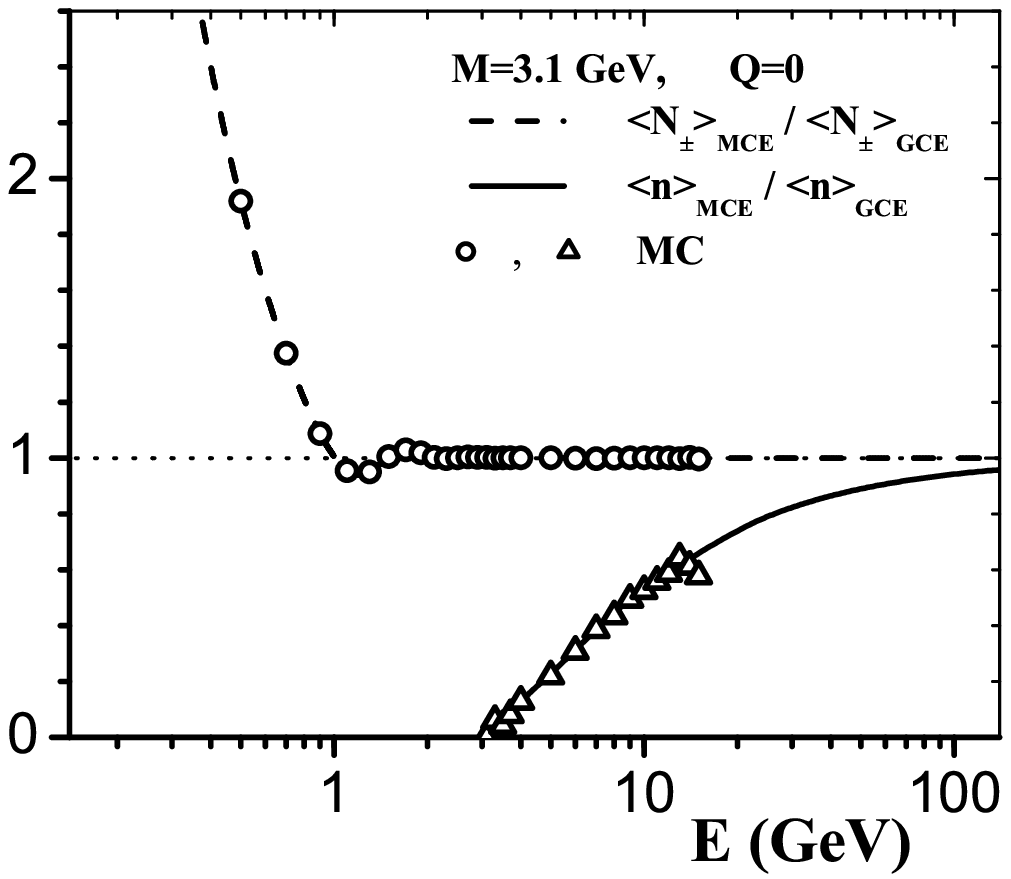,height=7cm,width=8cm}
 \caption{
The dependence of
the ratios of the mean  particle multiplicities calculated 
in the MCE to those the GCE on the total energy of the system.
The gas  with zero net charge $Q=0$ of
massless charged particles ($g_{\pm}=1$) and 
heavy neutral particles ($M = 3.1$~GeV and $G=1$) is
considered. The ratios for massless particles, 
$\langle N_{\pm}\rangle_{MCE}/\langle N_{\pm}\rangle_{GCE}$,
(dashed line) and heavy neutral particles,
$\langle n\rangle_{MCE}/\langle n\rangle_{GCE}$,
(solid line) are plotted.
The corresponding Monte-Carlo (MC) calculations are shown
by circles and triangles.
The calculations are performed for T~=~160~MeV and
$ E \equiv \langle E\rangle_{GCE}$ in the GCE and for
$ E \equiv \varepsilon(T) V$ in the MCE.
}\label{fig1}
\end{figure}

Now we calculate the mean particle multiplicities in the MCE and
compare them to the GCE multiplicities (\ref{GCE mult}). 
This comparison is performed at the same energy density which is
set to be the energy density calculated in the GCE for
$T=160$~MeV.  In the MCE
the volume $V$ 
equals to that in the GCE, and the MCE energy $E$ equals to GCE
average energy $E_{GCE}$ (\ref{GCE Energy}).
%
%\\
%\\
%
Let us start with the system of massless charged and heavy
neutral particles with zero net charge.  In this case all
MCE averages  are calculated as:
 \eq{ \label{<>mceQ0}
 \langle\ldots\rangle_{MCE}
 \,=\, \frac{1}{\Omega(E,\,V;\,Q=0,\,M)}
 \sum_{N_{\pm},n=0}^{\infty}\ldots \Omega_{N_{\pm},n}(E,\,V)\,,
 }
where $\Omega_{N_{\pm},n}(E,\,V)$ is given by Eq.~(\ref{WNpm,n(E,V)}),
$\Omega(E,V; Q=0,M)\equiv\sum_{N_{\pm},n=0}^{\infty}
\Omega_{N_{\pm},n}(E,\,V)$, and $\Omega_{0,0}(E,\,V)=0$,
$n_{max}=\left[\frac{E}{M}\right]$ due to the exact energy
conservation. The degeneracy factors are fixed to be $g_{\pm}=G=1$.
In Fig.~\ref{fig1} the ratios of average particle numbers in the MCE to
those in the GCE are presented as functions of the system energy
$E\equiv \langle E\rangle_{GCE}$. We remind that $V\equiv
E/\varepsilon(T)$. Circles and triangles represent Monte-Carlo
(MC) calculations  for the same parameters (see Section III for details
of the used procedure). These calculations are
in an  agreement with the analytical results. 
Further we also
show  the MC results obtained with the additional requirement of the exact
momentum conservation ($\bf P = \bf 0$) for the whole system.
For the heavy particles with mass $M=3.1$~GeV one observes 
(see  Fig.~\ref{fig1}) the MCE suppression of the heavy particle multiplicity.
This suppression is  strong at energies $E$ comparable to
heavy particle mass $M$, and it is still significant  at $E=100$ GeV
$\gg M$. 

In Fig.~\ref{fig2}~(left) the ratios $\langle
N_{\pm}\rangle_{MCE}/\langle N_{\pm}\rangle_{GCE}$ and
$\langle n\rangle_{MCE}/\langle n\rangle_{GCE}$ are shown
for a smaller value of the heavy particle mass, $M=0.7$~GeV. One
observes a fast drop of the $\langle
N_{\pm}\rangle_{MCE}/\langle N_{\pm}\rangle_{GCE}$ and
strong ``oscillations'' of the $\langle n\rangle_{MCE}/\langle
n\rangle_{GCE}$ ratio. The MCE  enhancement effect 
($\langle n\rangle_{MCE}/\langle n\rangle_{GCE}> 1$)  
increases strongly with decreasing mass $M$ of the heavy particle.
Note that the non-relativistic treatment of the heavy particles is
still approximately valid at $M=0.7$~GeV and $T=160$~MeV.

Let us consider the massless charged and heavy neutral
particles with the net charge $Q=2$. In this
case all MCE averages are also calculated by means of
Eq.~(\ref{<>mceQ0}). Besides, $\langle
N_+\rangle_{MCE}$ and $\langle N_-\rangle_{MCE}$ are
calculated independently using
Eqs.~(\ref{WN+,n(E,V,Q)},\ref{WN-,n(E,V,Q)}). The corresponding GCE
values now contain $\exp(\pm\mu/T)$, and the equation $\langle
N_+\rangle_{GCE}-\langle N_-\rangle_{GCE}=Q$ should be
solved:
 \eq{ \label{<N+>-<N->=Q(1)}
 \frac{g_{\pm}V\,T^3}{\pi^2}\,
 \left[\exp\left(\frac{\mu}{T}\right)~-~\exp\left(-\frac{\mu}{T}
 \right)\right]~
 =~Q~,
}
where the volume is given by Eq.~(\ref{GCE Energy}) as
$\langle E\rangle_{GCE}/\varepsilon(T)$. In order to make a simple
estimate, we  neglect a small  contribution of  heavy
particles to the energy density and consequently Eq.~(\ref{<N+>-<N->=Q(1)})
transforms into:
 \eq{
 \frac{E}{3T}\,\tanh\left(\frac{\mu}{T}\right)
% \frac{e^{+\frac{\mu}{T}}-e^{-\frac{\mu}{T}}}
%      {e^{+\frac{\mu}{T}}+e^{-\frac{\mu}{T}}}
\,=\,Q\,.
 }
This equation has  solutions
for $E > 3T\,Q$.
The results for the MCE to GCE particle ratios
are presented in
Fig.~\ref{fig2}~(right).
Negatively charged
particles are suppressed because of the exact charge conservation,
and the first maximum seen in
Fig.~\ref{fig2}~(left) for heavy neutral particles disappears.

\subsection{Canonical Ensemble}\label{sub-Sec-CE}

Let us consider the CE system with zero total net charge,
$Q=0$. A standard way is to compare the CE and GCE results at the
same volume $V$ and temperature $T$
(see e.g. \cite{ce-c,ce-d,ce-e}).
One observes here the reduction of charged particle
multiplicities with decreasing system volume in comparison
to ones calculated in the thermodynamical limit. 
The effect is called canonical suppression.
Note that the
energy density in the CE, $\varepsilon_{CE}$, is also suppressed
in  comparison with that in the GCE at the same temperature. This is
due to the suppression of charged particle multiplicities.
Our aim is to compare the particle multiplicities in different
ensembles at the same energy density.
Thus, we perform the comparison at 
the same volume and energy in all ensembles, so that the
energy densities are also equal to each other,
$\varepsilon_{CE}=\varepsilon_{MCE}=\varepsilon_{GCE}=E/V=const$.
Of course, under this condition the mean multiplicities and temperatures 
calculated in the CE
and the GCE are equal in the
thermodynamic limit, $V\rightarrow\infty$. The results are, however,
different for small systems. When the volume of the system in the CE
decreases, but the energy density is kept fixed,
the temperature increases which
leads to  an increase of the heavy neutral particle multiplicity.
The CE temperature, $T^*$, is obtained
as a solution of the equation:
 \eq{ \label{CEQ01}
\frac{\langle E\rangle_{CE}}{V}~\equiv~
 6T^* \frac{\langle N_{\pm}\rangle_{CE}}{V}
 \,+\, \left(\frac{3}{2} T^* + M\right)\, \frac{\langle
 n\rangle_{CE}}{V} \,=\, \frac{E}{V}~=~\varepsilon_{GCE} \,=\,
 const\, ,
 }
where ($z\equiv gV T^{*3}/\pi^2$)
 \eq{ \label{CEQ02}
 \langle N_{\pm}\rangle_{CE}
 \,=\,
 %\frac{gV T^{*3}}{\pi^2}\,
 z~\frac{I_1(2z)}{I_0(2z)}\,,
 \qquad
 \langle n\rangle_{CE}
 \,=\, GV ~\left(\frac{MT^*}{2\pi}\right)^{3/2}\,\exp\left(
 -\frac{M}{T^*}\right)~.
} A comparison of Eq.~(\ref{CEQ02}) and Eq.~(\ref{GCE mult})
demonstrates the CE suppression of charged multiplicities. This
leads to $T^*>T$ due to Eq.~(\ref{CEQ01}) hence to $\langle
n\rangle_{CE}>\langle n\rangle_{GCE}$~, as the equations for
neutral particle multiplicities have the same form in the GCE and the CE.
The results are shown in Fig.~\ref{fig3}. Massless particles are
suppressed in the CE, Fig.~\ref{fig3} (left), since the increase of
temperature is insufficient to overcome the CE suppression.
A smooth increase of the ratio
$\langle n \rangle_{CE} / \langle n \rangle_{GCE}$
is seen in Fig.~\ref{fig3} (right).
Thus,
a fast increase just above threshold and the following ``oscillations'' of the
$\langle n\rangle_{MCE}/\langle n\rangle_{GCE}$ ratio are due to
energy-momentum conservation laws. 

\subsection{ Grand Micro-Canonical Ensemble}

We turn now  to the discussion of 
the system with neutral particles only, both
massless and heavy ones. When the energy is exactly conserved we
call this ensemble the grand micro-canonical ensemble 
(GMCE) in order to
distinguish it from the MCE where both the energy and the charge are
strictly conserved. 
The calculation within
the GMCE allows us to check 
whether the MCE
enhancement is only related to the  charge conservation or 
it is due to the
energy conservation.
We set
$g_{\pm}=0$ and $g=G=1$, and in this case all GMCE averages are
calculated as:
 \eq{ \label{<>mce}
\langle\ldots\rangle_{GMCE} \,=\, \frac{1}{\Omega(E,\,V;\,M)}
 \sum_{N,n=0}^{\infty} \ldots \Omega_{n,N}(E,\,V)\,,
 }
where $\Omega(E,\,V;\,M)\equiv \sum_{N,n=0}^{\infty} \Omega_{n,N}(E,\,V)$ is
the total MCE partition function, where $\Omega_{0,0}(E,\,V)=0$ and
$n_{max}=\left[\frac{E}{M}\right]$ due to  the exact energy
conservation. The result is shown in Fig.~\ref{fig4}, where one can
see that the heavy particle enhancement near the threshold is 
also present in the
neutral system. The amplitude of a drop in $\langle
N\rangle_{GMCE}/\langle N\rangle_{GCE}$ and the ``oscillations''
of $\langle n\rangle_{GMCE}/\langle n\rangle_{GCE}$ becomes
smaller than in the MCE with charged massless particles, but they
are still present. The Monte-Carlo (MC) calculations with the total
momentum conservation show a similar behavior: a decrease of
the ratio for the massless particles and an increase for the heavy
particles. The momentum conservation leads to an increase of both
ratios, it also smears and
shifts the peak for heavy particles.
In Fig.~\ref{fig5} we present the energy dependence of  the particle multiplicities
calculated in the GMCE and
the GCE. 
%massless particle multiplicities, $\langle
%N\rangle_{g.MCE}$,
%,\, \langle N\rangle_{MCE}$ --
%and heavy
%particle multiplicities, $\langle n\rangle_{g.MCE}$,
%,\, \langlen\rangle_{MCE}$ --
%for $M=0.7$~GeV. 
%The momentum conservation leads to an increase of
%average values in the GMCE, because it demands at least two particles
%in the final state.
%

%*********************************************************************************
\section{The hadron-resonance gas model
}
\label{microca}
%*********************************************************************************

This section starts with a brief presentation of the numerical methods used to
calculate mean hadron multiplicities in the MCE for
the ideal hadron-resonance gas. The full description can be found in 
\cite{shahme1,shahme2}.
Further on, we  show and discuss the results on the energy dependence of hadron yields
near the threshold.  

\subsection{The Procedure}

The MCE partition function is defined as the sum over
all multi-hadronic states $| h_V \rangle$ localized within the
volume $V$ of the system and constrained with the four-momentum and
the abelian (i.e. additive) charge conservation:
\begin{equation}\label{micro1}
 \Omega = \sum_{h_V} \langle h_V | \delta^4 (P- P_{\rm op})
 \delta_{\Qz,\Qz_{\rm op}} | h_V \rangle\, ,
\end{equation}
where $\Qz = (Q_1,\ldots,Q_M)$ is a vector of $M$ integer abelian
charges (electric, baryon number, strangeness etc.), $P$ is the
overall four-momentum of the system and $P_{\rm op}$, $\Qz_{\rm
op}$ the relevant operators. Provided that relativistic quantum
field effects are neglected and the volume of the system is large
enough to allow the approximation of finite-volume Fourier
integrals with Dirac deltas, it can be proved \cite{shahme1} that
the micro-canonical partition function $\Omega$ can be written as a
multiple integral:
\begin{eqnarray}\label{micro2}
\Omega= \frac{1}{(2\pi)^{4+M}} \int \d^4 y \,\e^{\i P \cdot y}
\int_{-\pivs}^{+\pivs} \d^M \phi \, \e^{\i \Qz \cdot \phivs}
\exp\Big[\sum_j \frac{(2J_j+1)V}{(2\pi)^3} \!\!\!
 \int \d^3\p \, \log (1\pm \e^{-\i p_j \cdot y -\i \qj \cdot \phivs})^{\pm 1}\Big]\, ,
\end{eqnarray}
where $\qj$ is the vector of the abelian charges for the $j^{\rm
th}$ hadron species, $J_j$ its spin and $p_j=(\sqrt{m^2_j+\p^2},{\bf p})$, 
where $m_j$ is the mass of the $j^{\rm
th}$ hadron species; the upper sign applies to
fermions, the lower to bosons. The integral (\ref{micro2}) is more
easily calculable in the rest frame of the system where $\bf P=\bf 0$. 
Unfortunately, an analytical solution with no charge
constraint is known only in two limiting cases treated in previous
sections: non-relativistic and ultra-relativistic (i.e. with all
particle masses set to zero). The full relativistic case has been
attacked with several kinds of expansions \cite{cerhag1} but none
of them proved to be fully satisfactory as the achieved accuracy
in the estimation of different kinds of averages could vary from
some percent to a factor 10. Therefore, a numerical integration of
Eq.~(\ref{micro2}) is needed. The most suitable method is to
decompose $\Omega$ into the sum of the phase space volumes with
fixed particle multiplicities for each species:
\begin{equation}\label{microdec}
  \Omega = \sum_{\Nj} \Omega_{\Nj} \delq\, ,
\end{equation}
$\Nj$ being a vector of $K$ integer numbers $(N_1,\ldots,N_K)$, i.e. the multiplicities
of all of the $K$ hadronic species.

Here we use the approximated expression
in Boltzmann statistic of the phase space volume $\Omj$ for the channel $\Nj$
in order to make a qualitative study on
micro-canonical effects in a coherent way with the analysis in the
analytical model made in the previous part of the paper. For the
same reason, we will keep fixed masses for hadron resonances
neglecting the Breit Wigner broadening effect. Both of these
simplifications, do not affect the results significantly.
The phase
space volume for fixed multiplicities in Boltzmann statistic reads \cite{shahme1}:
\begin{eqnarray}\label{boltz}
\Omj = \prod_j \frac{V^{N_j}(2J_j+1)^{N_j}}{(2\pi)^{3N_j} N_j!}
  \int \d^3 \p_1 \ldots \d^3 \p_N \, \delta^4 (P-\sum_{i=1}^N p_i)\, ,
\end{eqnarray}
where $N = \sum_j N_j$.

We are mainly interested in the calculation of
quantities relevant to particle multiplicities, not to their
momenta, namely their kinematical state. The average of an
observable $O$ depending on particle multiplicities in the
micro-canonical ensemble can then be written as:
\begin{equation}\label{obs}
  \langle O \rangle = \frac{\sum_{\Nj} O(\Nj) \Omj \delq}
  {\sum_{\Nj} \Omj \delq}\, ,
\end{equation}
where $O(\Nj)\equiv N_k$ for the  average multiplicity of the
$k-$th hadron species $\langle N_k \rangle$. Altogether, what we
need to calculate in order to evaluate an average~(\ref{obs}) are
integrals like~(\ref{boltz}).

In order to
be able to effectively calculate $\Omj$ for any channel, a
brute force option is to do it for all of them. However, this
method is not appropriate for a system like the hadron gas,
as the actual number of channels is very large. Indeed, with 265
light-flavored hadrons and resonances (those included in the
latest Particle Data Book issue \cite{pdg}), the number of
channels allowed by the energy-momentum conservation is enormous and
increases almost exponentially with the cluster mass, which  
requires an unacceptably large computing time. 
Therefore, the calculation of
the phase space volume of all allowed channels is only possible 
for very light clusters, in practice lighter than $\sim 2$ GeV.
Hence, if a method based on the exhaustive exploration of the
channel space is not affordable, one has to resort to Monte-Carlo
methods, whereby the channel space is randomly sampled.

An estimate of the average (\ref{obs}) can be made by means of the
so-called importance sampling method. The idea of this method is
to sample the channel space (i.e. the set of integers $N_j$, one
for each hadron species) not uniformly, but according to an
auxiliary distribution $\Pij$ which must be suitable to being
sampled very efficiently to keep computing time low, and, at the
same time, as similar as possible to the distribution $\Omj$. The
latter requirement is dictated by the fact that $\Omj$ is sizable
over a very small portion of the whole channel space. Thus, if
random configurations were generated uniformly, for almost all of
them $\Omj$ would have a negligible value, and a huge number of
samples would be required to achieve a good accuracy. On the other
hand, if samples are drawn according to a distribution similar to
$\Omj$, little time is wasted to explore unimportant regions and
the estimation of the average (\ref{obs}) is more accurate. A
crucial requirement for $\Pij$ is not to be vanishing or far
smaller than $\Omj$ anywhere in its domain in order not to exclude
some good regions from being sampled, thereby biasing the
calculated averages in a finite statistics calculation.

A Monte-Carlo estimate of $\langle O \rangle$ is:
\begin{equation}\label{estimator}
  \langle O \rangle \doteq \frac{\sum_{k=1}^{N_S} O(\Nj^{(k)})
  {\displaystyle \frac{\Omj^{(k)}}{\Pij^{(k)}}} }
  {\sum_{k=1}^{N_S} {\displaystyle \frac{\Omj^{(k)}}{\Pij^{(k)}}} }\, ,
\end{equation}
where $\Nj^{(k)}$ are samples of the channel space extracted according to the
distribution $\Pi$ and fulfilling the charge constraint $\Qz= \sum_j N_j \qj$.

Provided that $N_S$ is large enough so that the distributions of
both numerator and denominator in Eq.~(\ref{estimator}) are
Gaussians (hence the conditions of validity of the central limit
theorem are met), the statistical error $\sigma_\Oav$ on the
average $\Oav$ can be estimated as:
\begin{eqnarray}\label{error}
\sigma_{\Oav}^2 = \frac{1}{N_S \Omega^2} \Bigg\{ {\sf E}_\Pi \left(
O^2 \frac{\Omj^2}{\Pij^2} \right) + \Oav^2  {\sf E}_\Pi \left(
\frac{\Omj^2}{\Pij^2} \right) - 2 \Oav   {\sf E}_\Pi \left( O
\frac{\Omj^2}{\Pij^2} \right) \Bigg\}\, ,
\end{eqnarray}
where ${\sf E}_\Pi$ stands for the expectation value relevant to the $\Pi$
distribution.

Here we  define  $\Pij$ as the product of $K$ (as many as particle species)
Poisson distributions:
\begin{equation}\label{poisson}
  \Pij = \prod_{j=1}^K \exp[-\nu_j ] \frac{\nu_j^{N_j}}{N_j!}
\end{equation}
which will be henceforth referred to as the {\em multi-Poisson
distribution} or MPD, enforcing as mean values the mean hadronic
multiplicities $\nu_j$ calculated in the GCE
with volume and mean energy equal to the volume and mass of the
system:
\begin{equation}\label{canmean}
  \nu_j = \frac{(2J_j+1) V}{2\pi^2} \, m_j^2 T {\rm K}_2 \left(\frac{m_j}{T}\right)
  \prod_{i} \lambda_i^{q_{ji}}\, ,
\end{equation}
where $V$ is the volume of the system, $T$ is the temperature and $\lambda_i$ the
fugacity corresponding to the charge $Q_i$. Temperature and fugacities are determined
by enforcing the GCE mean energy and charges to be equal to the actual
energy $E$ and charges $\Qz$ of the system:
\begin{eqnarray}\label{saddle}
 &&  E = T^2 \frac{\partial}{\partial T} \sum_j z_j(T) \prod_{i} \lambda_i^{q_{ji}}\, ,
 \nonumber \\
 &&  \Qz = \sum_j \qj z_j(T) \prod_{i}  \lambda_i^{q_{ji}}\, ,
\end{eqnarray}
with
\begin{equation}\label{zeta}
  z_j(T) = \frac{(2J_j+1) V}{2\pi^2} \, m_j^2 T {\rm K}_2 \left(\frac{m_j}{T}\right)\, .
\end{equation}
The distribution (\ref{poisson}) can indeed be sampled very efficiently and is
the actual multi-species multiplicity distribution in the GCE
 in the limit of Boltzmann statistics.

%**********************************************************************
\subsection{Threshold Effects in the Hadron-Resonance Gas}
%**********************************************************************

We have calculated the ratios of mean multiplicities of several
hadron species in the MCE and the CE with respect to the GCE for an ideal
hadron-resonance gas including all species up to a mass of about
1.8 GeV. Quantum statistic effects and resonance mass broadening
have been turned off. In order to make a proper comparison between the
different ensembles, energy (in the rest frame) and volume in both
the CE and the GCE have been set to the same value as in the MCE. This
implies that the energy density $\varepsilon$ has the same value
in all ensembles as well. It has been fixed to 0.3895 GeV/fm$^3$,
corresponding to a temperature of 160 MeV in the GCE. As it has
been discussed in Section II, in the CE, owing to
the exact charge conservation, the energy density is not an
intensive quantity but it also depends on the volume. This implies
that for a fixed energy density, temperature varies as a function
of total energy. In particular, for a completely neutral hadron
gas which will be discussed henceforth temperature decreases as
energy increases. This happens because at lower energies or
volumes charge conservation suppresses more and more charged
particle multiplicities and the system needs an increase in
temperature to keep the energy density constant.

The results of our calculations are shown in
Figs.~\ref{fig6}-\ref{fig13}, as a function of the total energy
near the production threshold. In general, it can be seen from
these plots that for an actual ideal hadron-resonance gas the
(micro-)canonical enhancement for some particle species near the
threshold (which was implicitly numerically observed in
\cite{shahme2}) shows the same qualitative features as those of the
analytical model. Nevertheless, the correlation effects
between different particle species due to charge and
the energy-momentum conservation play an important role from the
quantitative point of view and results in abrupt changes in the
micro-canonical mean multiplicities as the total energy exceeds the
threshold for the production of a new channel.

In Fig.~\ref{fig6} we present the ratio between the average
multiplicities of $\pi^0$-mesons in the MCE (CE) and in the GCE as 
closed circles
(dashed line). Since pions are the lightest hadrons, we expect
them to play the same role as massless particles in the analytical model.
The ratios CE/GCE and MCE/GCE approach each other at energy of
about 2 GeV and, thereafter, slowly converge to the GCE limit. For
lower energies (less than $\sim 2$ GeV) the total particle
multiplicity in the MCE is $\sim 2$, as the dominant channels are
$\pi^0+\pi^0$ and $\pi^++\pi^-$. Therefore, while in the GCE the
mean multiplicity of $\pi^0$-mesons is proportional to the energy, in
the MCE it is almost constant and $\sim 2/3$. This explains why
the ratio $\langle \pi^0\rangle_{\rm MCE}/\langle\pi^0\rangle_{\rm GCE}$ has a nearly
hyperbolic shape at small energies.

On the other hand, the increase of the ratio $\langle \pi^0 \rangle_{\rm
CE}/\langle \pi^0 \rangle_{\rm GCE}$ as energy decreases is due to the higher
temperature in the canonical ensemble for small systems with the same
energy density (see Section~\ref{sub-Sec-CE} and the 
discussion above). As the heavy (i.e. with $m \gg T$) neutral particle
multiplicity is proportional to $\exp[-m/T]$ according to
Eq.~(\ref{CEQ02}), the ratio $\langle n_h \rangle_{CE}/\langle
n_h \rangle_{GCE}$ turns out to be proportional to:
\begin{equation}\label{ratio}
  \frac{\langle n_h \rangle_{CE}}{\langle n_h \rangle_{GCE}}
  \sim \exp{ \left[ -m_h \left( \frac{1}{T(V,\varepsilon)}-
  \frac{1}{T(\infty, \varepsilon)} \right) \right] }
\end{equation}
which grows exponentially with the particle mass. Indeed, in
Figs.~\ref{fig7}-\ref{fig10} the ratios for heavy neutral particles
$\eta$, $\rho$, $f_2$ and $J/\psi$ show the expected behavior,
with a ``canonical'' enhancement larger for heavier particles. For
$J/\psi$, it is so large that it exceeds the ``micro-canonical''
enhancement, which, conversely, tends to diminish at larger masses.
Indeed, for a sufficiently large mass, this enhancement disappears
as it should have been expected from the result of the analytical model
discussed in Section~\ref{sec2}.

\begin{figure}[h!]
 \epsfig{file=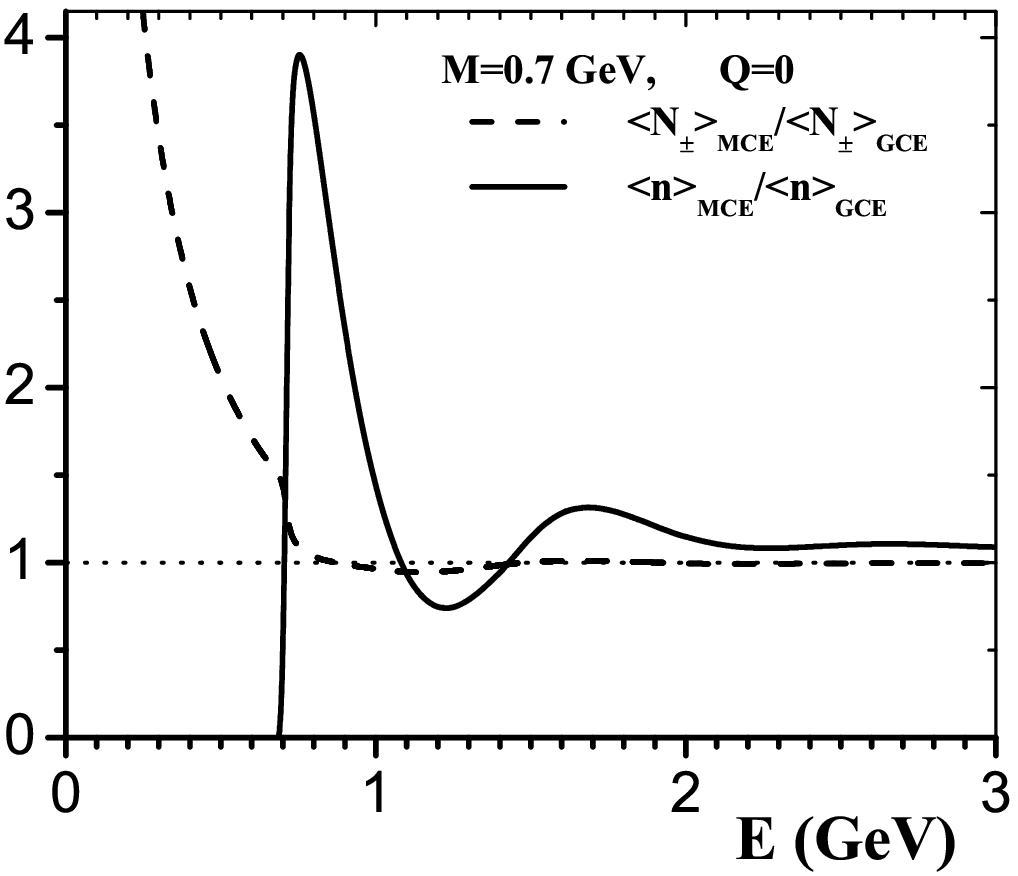,height=7cm,width=8cm}
 \hspace{0.5cm}
 \epsfig{file=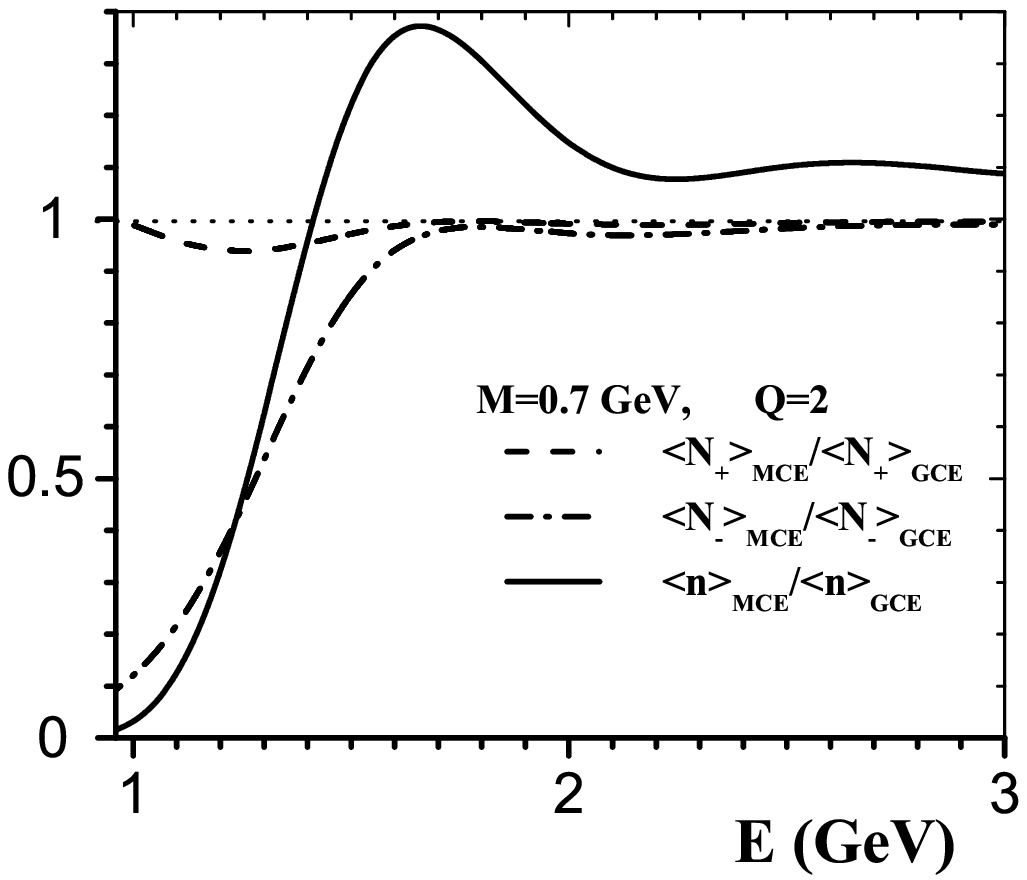,height=7cm,width=8cm}
 \caption{Left: 
The same as in Fig.~\ref{fig1} but for heavy particle mass
$M = 0.7$ GeV.
 \\
Right: 
The same as in Fig.~\ref{fig1} but for heavy particle mass
$M = 0.7$ GeV and the system net charge $Q=2$.
}\label{fig2}
\end{figure}

\begin{figure}[h!]
 \epsfig{file=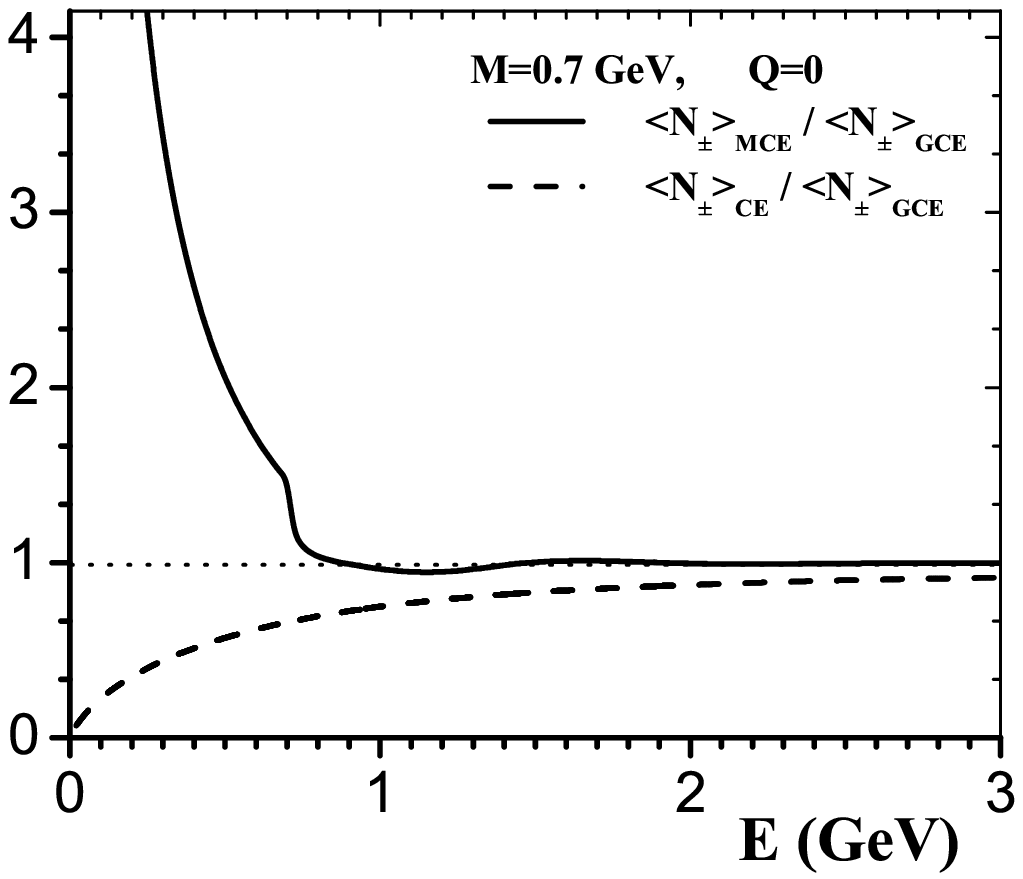,height=7cm,width=8cm}
 \hspace{0.5cm}
 \epsfig{file=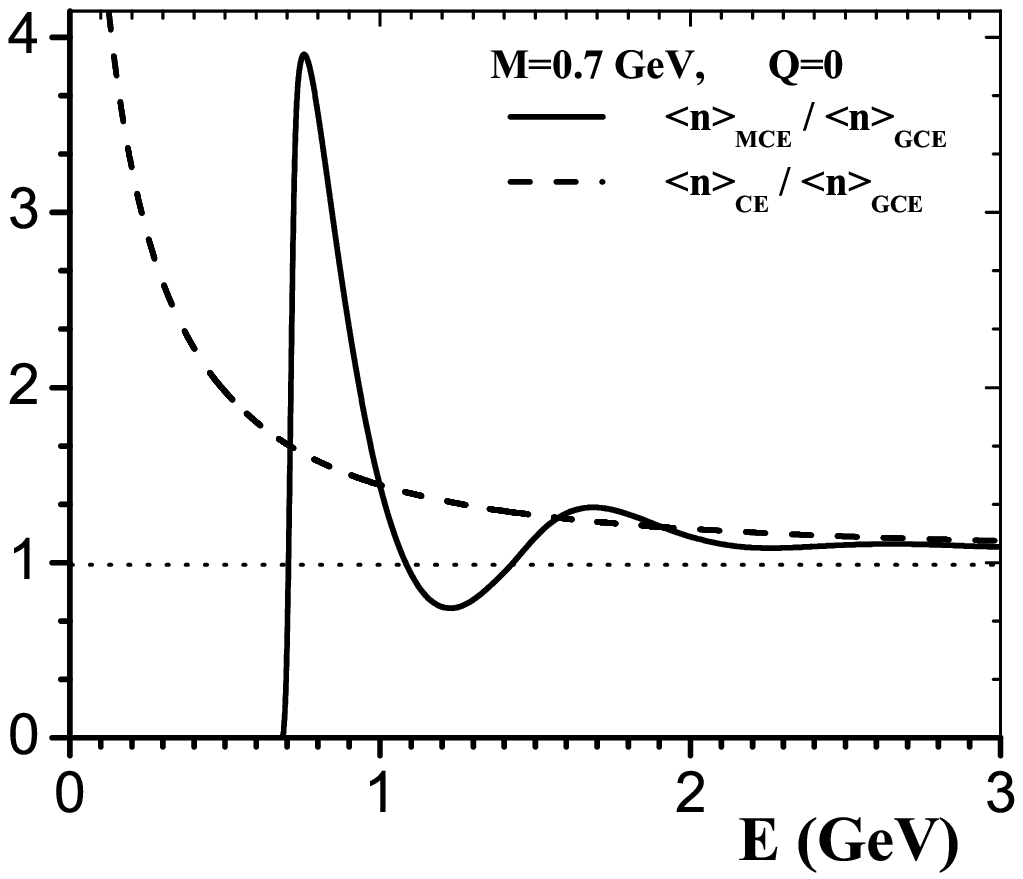,height=7cm,width=8cm}
 \caption{
The dependence of
the ratios of the mean  particle multiplicities calculated 
in the MCE and the CE to those in the GCE on the total energy of the system.
The gas  with zero net charge $Q=0$ of
massless charged particles ($g_{\pm}=1$) and 
heavy neutral particles ($M = 0.7$~GeV and $G=1$) is
considered. The ratios for massless particles, 
$\langle N_{\pm}\rangle_{MCE}/\langle N_{\pm}\rangle_{GCE}$ 
(solid line) and
$\langle N_{\pm}\rangle_{CE}/\langle N_{\pm}\rangle_{GCE}$
(dashed line) are shown in the left plot whereas
the ratios for heavy particles,
$\langle n_{\pm}\rangle_{MCE}/\langle n_{\pm}\rangle_{GCE}$
(solid line) and
$\langle n_{\pm}\rangle_{CE}/\langle n_{\pm}\rangle_{GCE}$
(dashed line) are shown in the right plot.
The calculations are performed for T~=~160~MeV and
$ E \equiv \langle E\rangle_{GCE}$ in the GCE and for
$ E \equiv \varepsilon(T) V$ in the MCE and in the CE.
}\label{fig3}
\end{figure}

\begin{figure}[h!]
 \epsfig{file=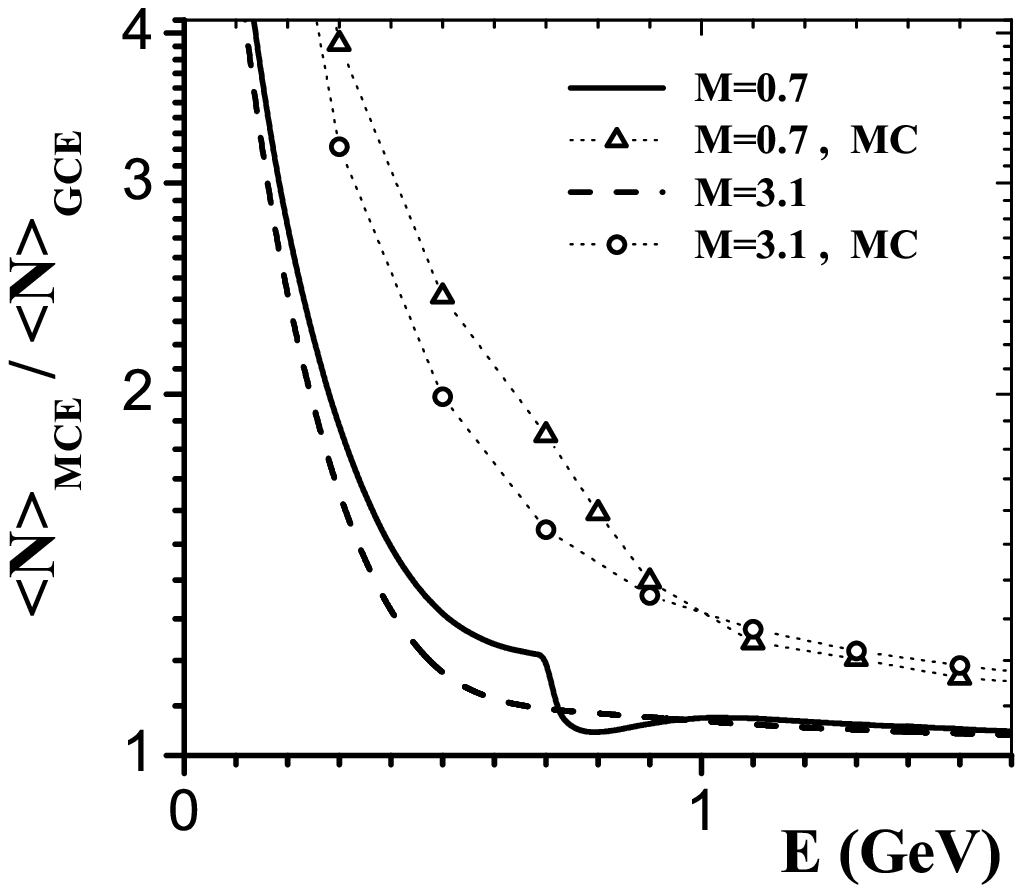,height=7cm,width=8cm}
 \hspace{0.5cm}
 \epsfig{file=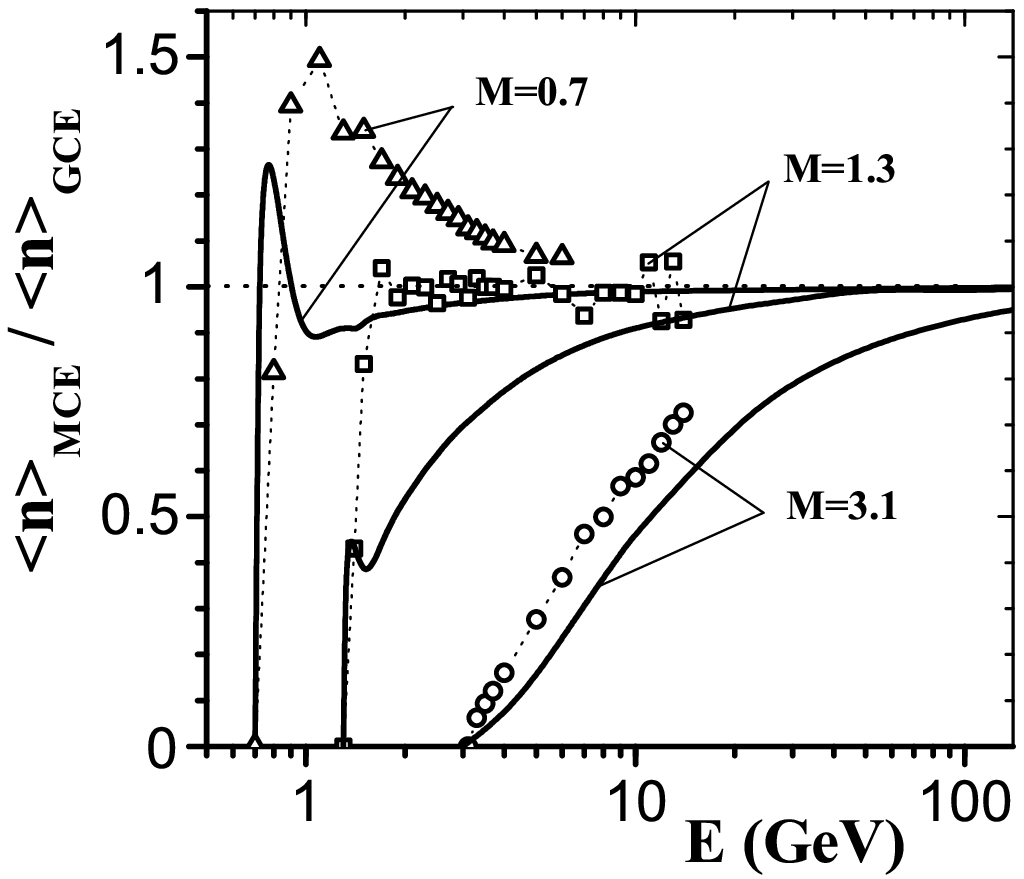,height=7cm,width=8cm}
 \caption{
The dependence of
the ratios of the mean  particle multiplicities calculated 
in the MCE to those in the GCE on the total energy of the system.
The gas  of
massless neutral particles ($g_{\pm}=1$) and 
heavy neutral particles ($M = 0.7,~1.3,~3.1$~GeV and $G=1$) is
considered. Left: The ratios for massless particles, 
$\langle N_{\pm}\rangle_{MCE}/\langle N_{\pm}\rangle_{GCE}$ 
calculated within the analytical model without momentum
conservation are shown by solid ($M = 0.7$ GeV) and
dashed ($M = 3.1$ GeV) lines. The corresponding Monte-Carlo
results with momentum conservation are indicated
by triangles and circles, respectively.
Right: The ratios for heavy particles,
$\langle n_{\pm}\rangle_{MCE}/\langle n_{\pm}\rangle_{GCE}$ 
calculated within the analytical model without momentum
conservation are shown by solid 
lines. The corresponding Monte-Carlo
results with momentum conservation are indicated
by triangles, squares and circles.
The calculations are performed for T~=~160~MeV and
$ E \equiv \langle E\rangle_{GCE}$ in the GCE and for
$ E \equiv \varepsilon(T) V$ in the MCE.
Connecting lines between the MC-dots are drawn to guide the eyes.}
\label{fig4}
\end{figure}

\begin{figure}[h!]
 \includegraphics[width=0.7\textwidth]{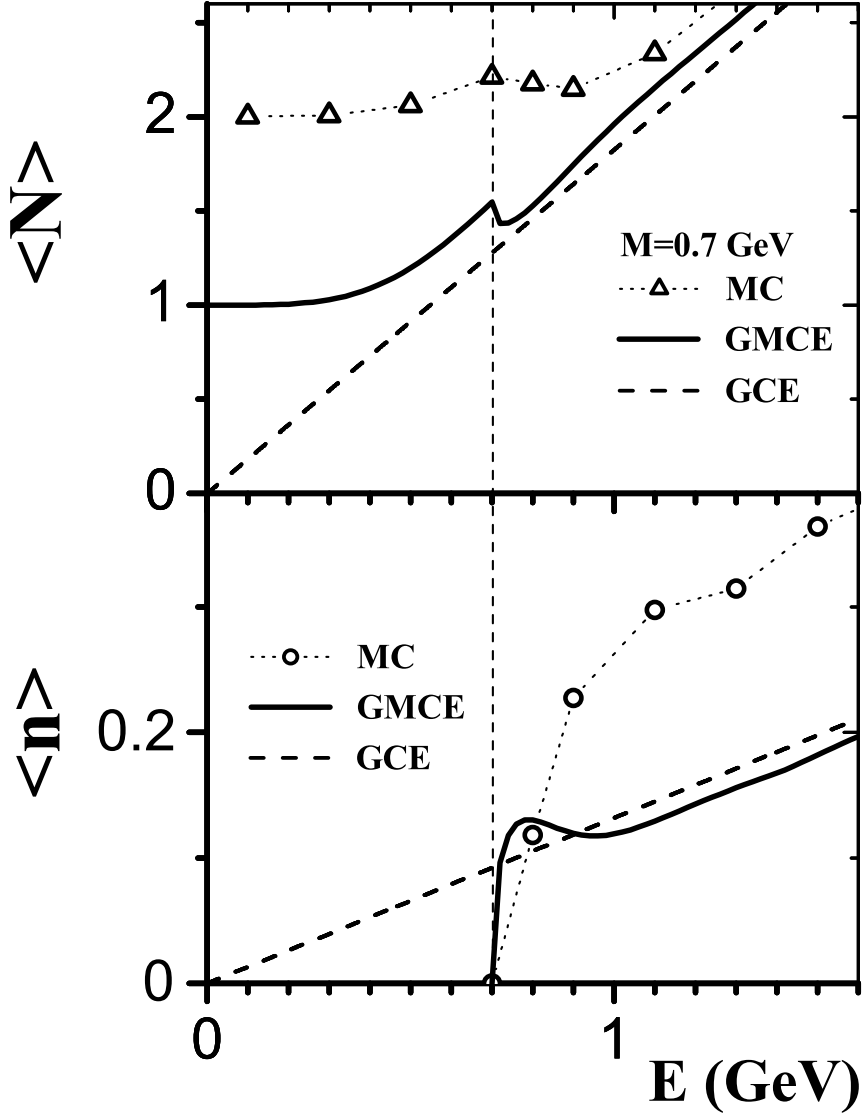}
 \caption{The dependence of
the mean  particle multiplicities calculated 
in the MCE, the GMCE and the GCE on the total energy of the system.
The gas  of
massless neutral particles ($g_{\pm}=1$) and 
heavy neutral particles ($M = 0.7$~GeV and $G=1$) is
considered. Top panel shows the multiplicity of massless particles, 
$\langle N_{\pm}\rangle$, 
calculated in the GCE (dashed line), in the GMCE (solid line)
and in the MCE with momentum conservation (triangles). 
Bottom panel shows the corresponding results for the heavy particles.
The calculations are performed for T~=~160~MeV and
$ E \equiv \langle E\rangle_{GCE}$ in the GCE and for
$ E \equiv \varepsilon(T) V$ in the MCE.
Connecting lines between the MC-dots are drawn to guide the eyes.
}\label{fig5}
\end{figure}

\begin{figure}[h!]
 \includegraphics[width=1.0\textwidth]{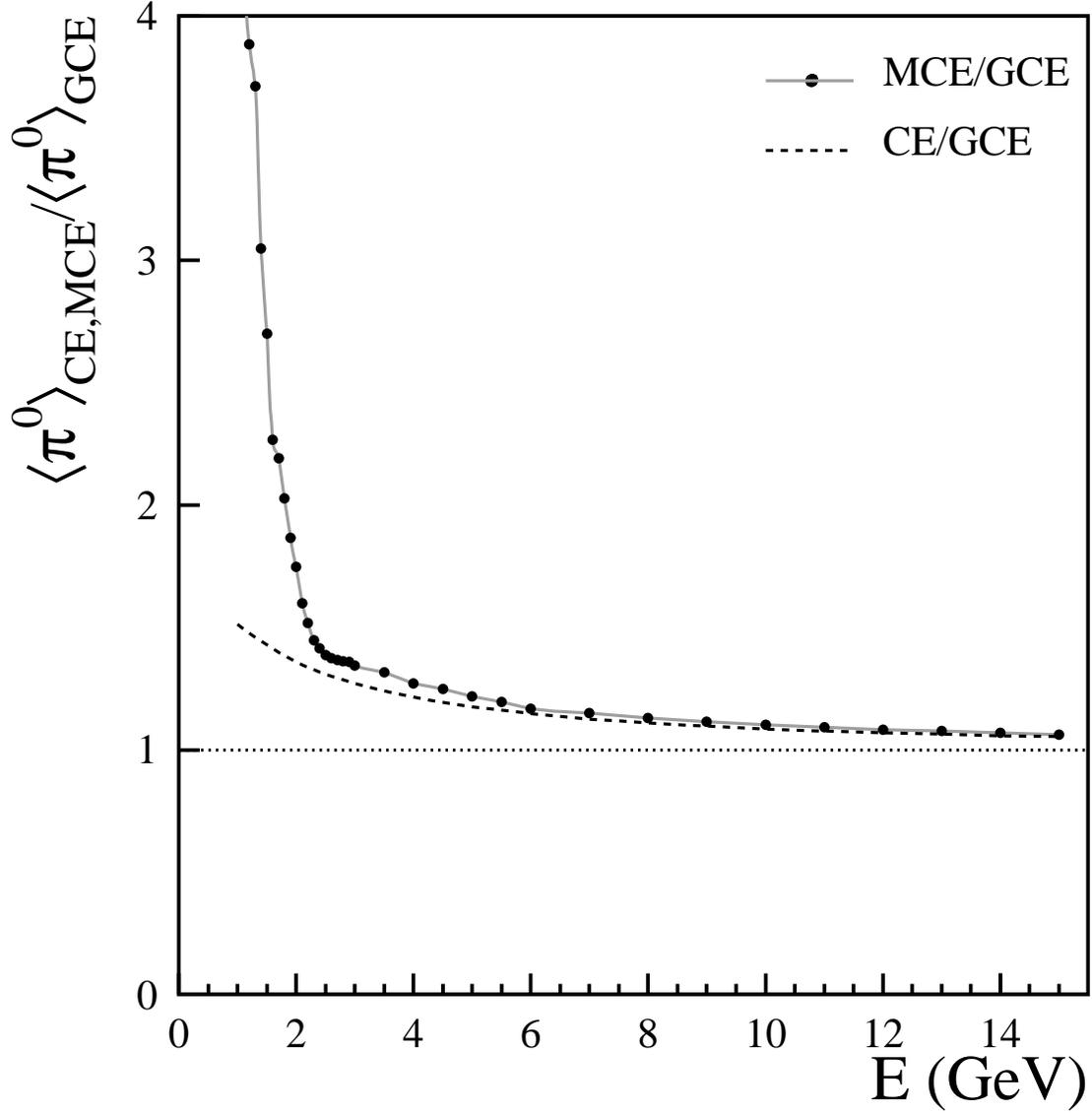}
\caption{
The energy dependence of the ratio of
the mean  $\pi^0$-meson multiplicity
in the MCE and the CE to those in the GCE obtained within
the hadron-resonance gas model.
The calculations are performed for T~=~160~MeV and
$ E \equiv \langle E\rangle_{GCE}$ in the GCE and for
$ E \equiv \varepsilon(T) V$ in the MCE and the CE.
Connecting lines between the MC-dots are drawn to guide the eyes.
}\label{fig6}
\end{figure}

\begin{figure}[h!]
 \includegraphics[width=1.0\textwidth]{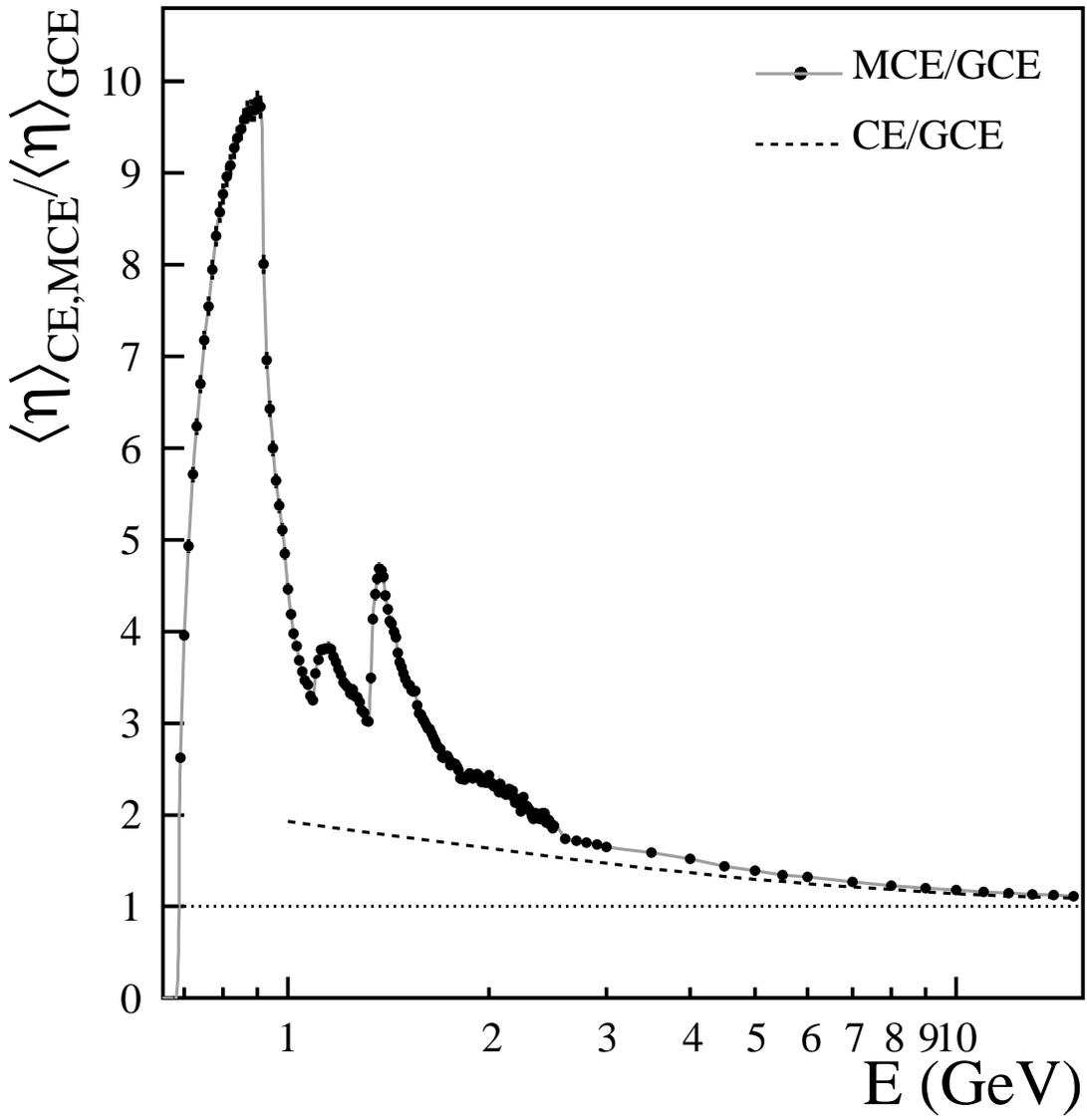}
\caption{
The same as in Fig.~\ref{fig6} but for $\eta$ meson.
}\label{fig7}
\end{figure}

\begin{figure}[h!]
 \includegraphics[width=1.0\textwidth]{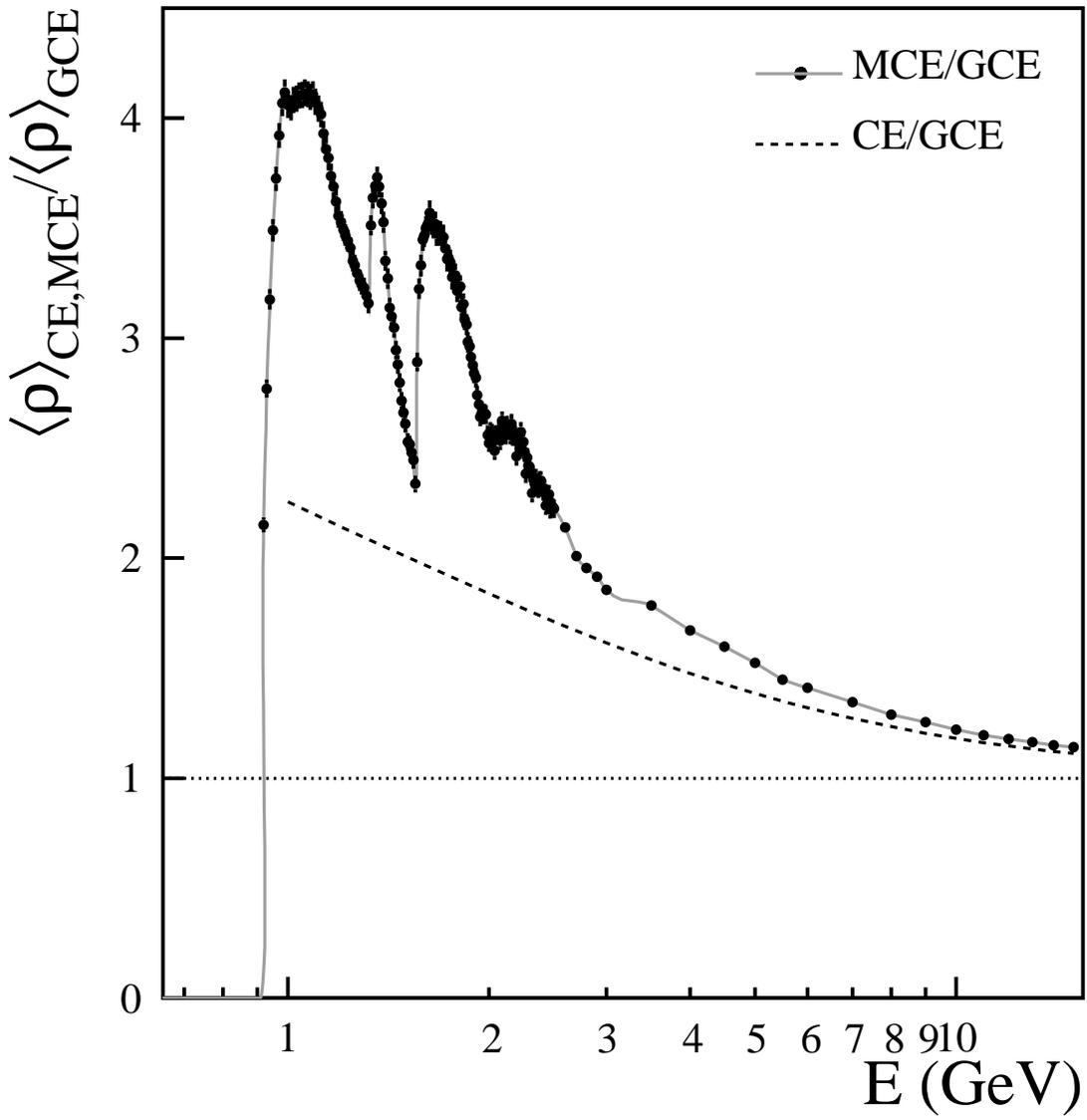}
\caption{
The same as in Fig.~\ref{fig6} but for $\rho$ meson.
 }\label{fig8}
\end{figure}

\begin{figure}[h!]
 \includegraphics[width=1.0\textwidth]{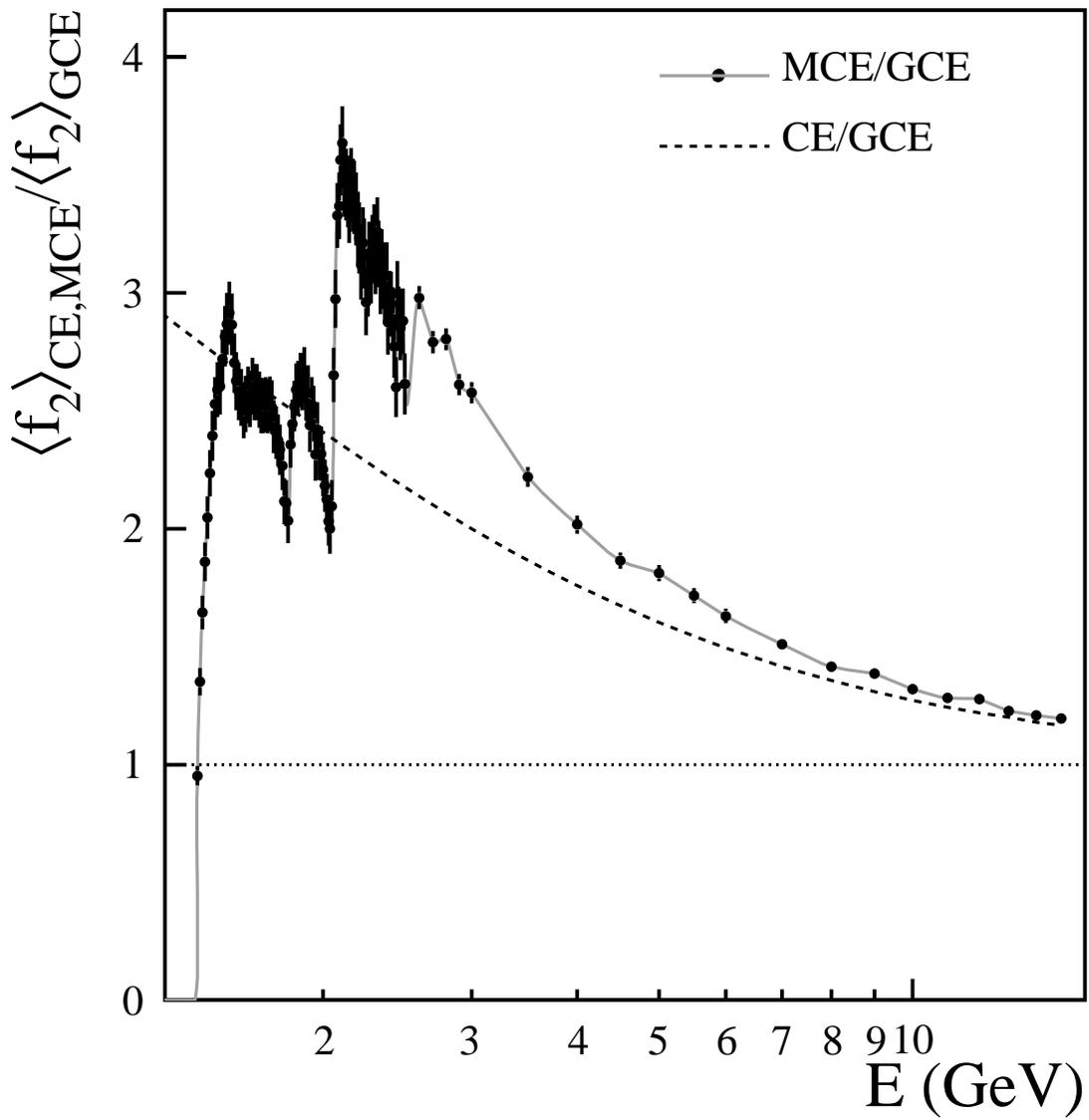}
 \caption{
The same as in Fig.~\ref{fig6} but for $f_2$ meson.
 }\label{fig9}
\end{figure}

\begin{figure}[h!]
 \includegraphics[width=1.0\textwidth]{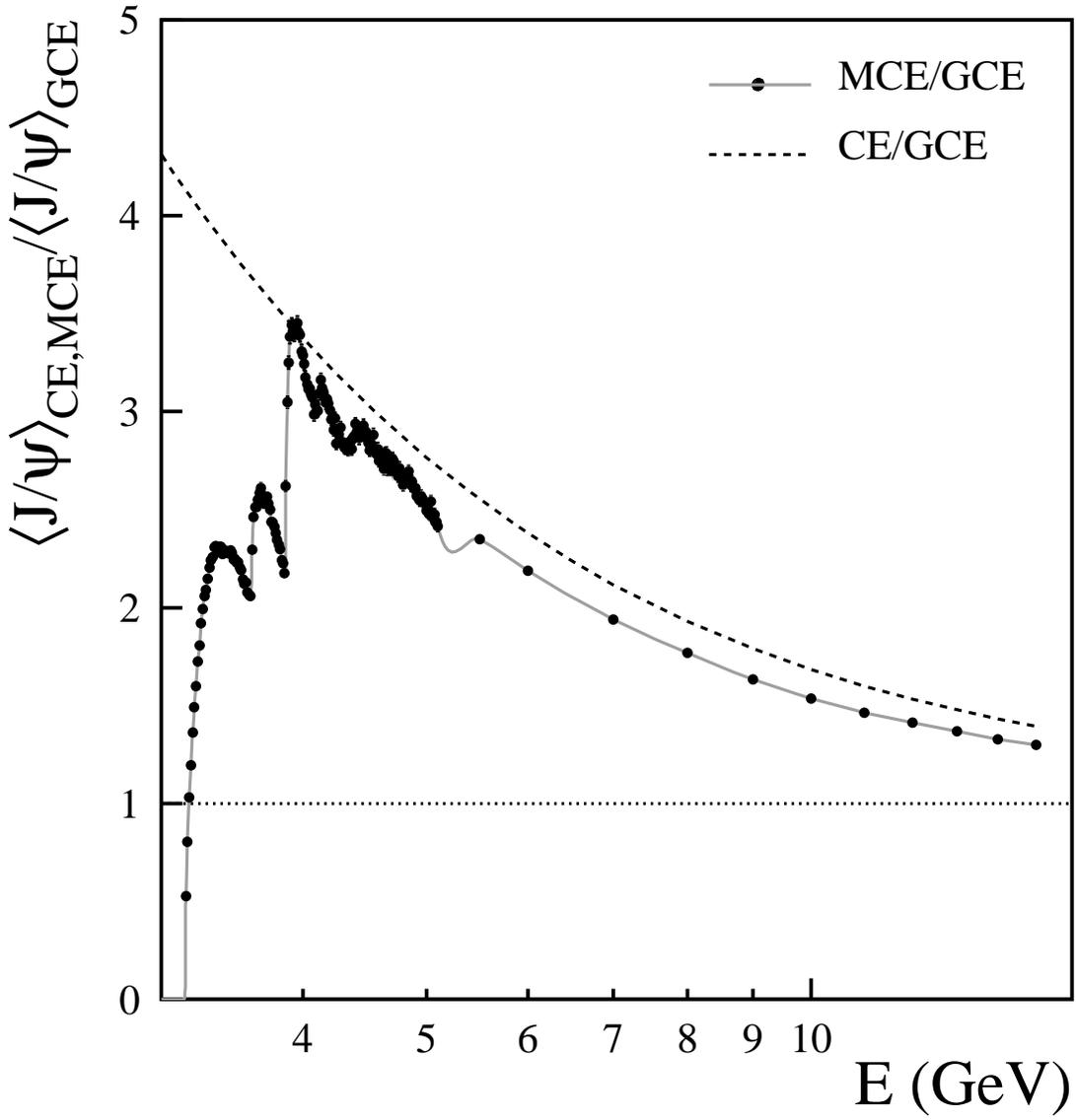}
\caption{
The same as in Fig.~\ref{fig6} but for $J/\psi$ meson.
}\label{fig10}
\end{figure}

\begin{figure}[h!]
\includegraphics[width=0.7\textwidth]{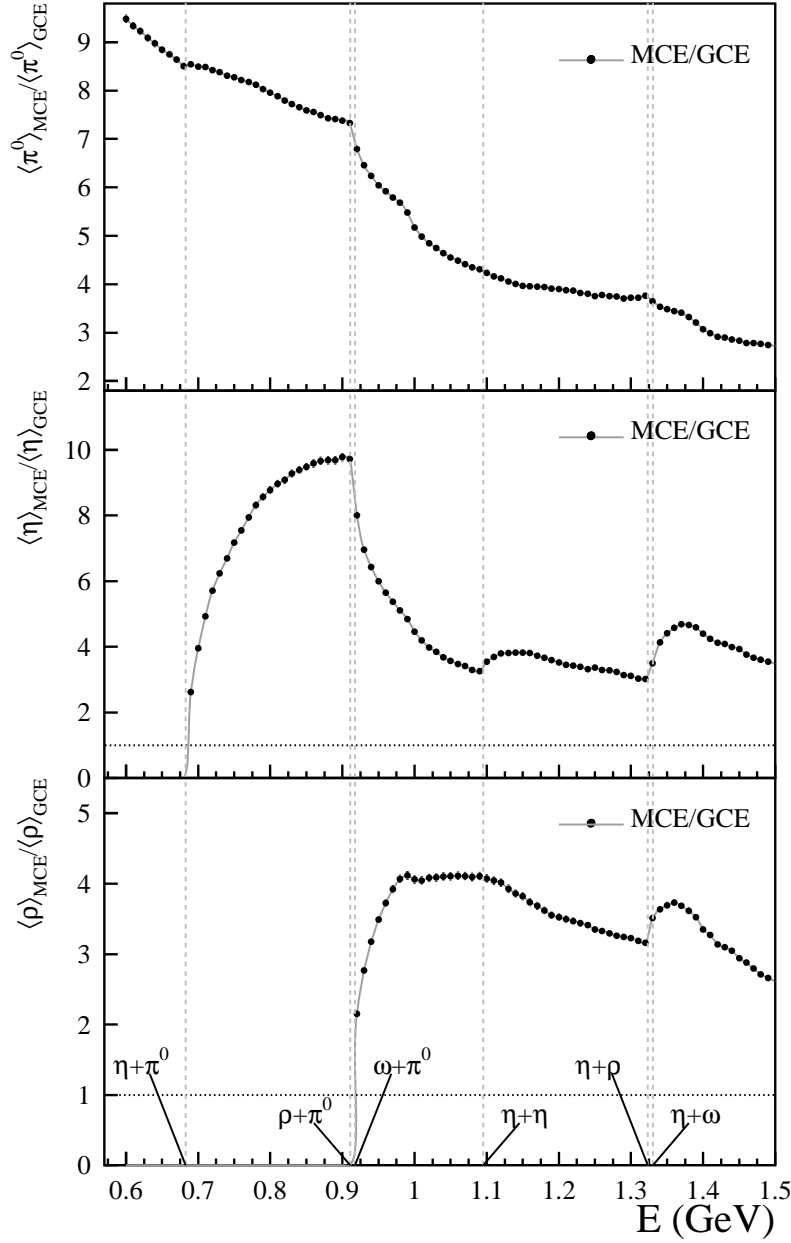}
\caption{
The energy dependence of the ratio of
the mean  multiplicities of $\pi^0$ (top),
$\eta$ (middle) and $\rho$ (bottom) mesons
in the MCE and the CE to those in the GCE obtained within
the hadron-resonance gas model.
The vertical dashed lines indicates  the threshold energies
of several two particle channels which are responsible for the
observed structures.
The calculations are performed for T~=~160~MeV and
$ E \equiv \langle E\rangle_{GCE}$ in the GCE and for
$ E \equiv \varepsilon(T) V$ in the MCE.
Connecting lines between the MC-dots are drawn to guide the eyes.
}\label{fig11}
\end{figure}

\begin{figure}[h!]
\includegraphics[width=0.8\textwidth]{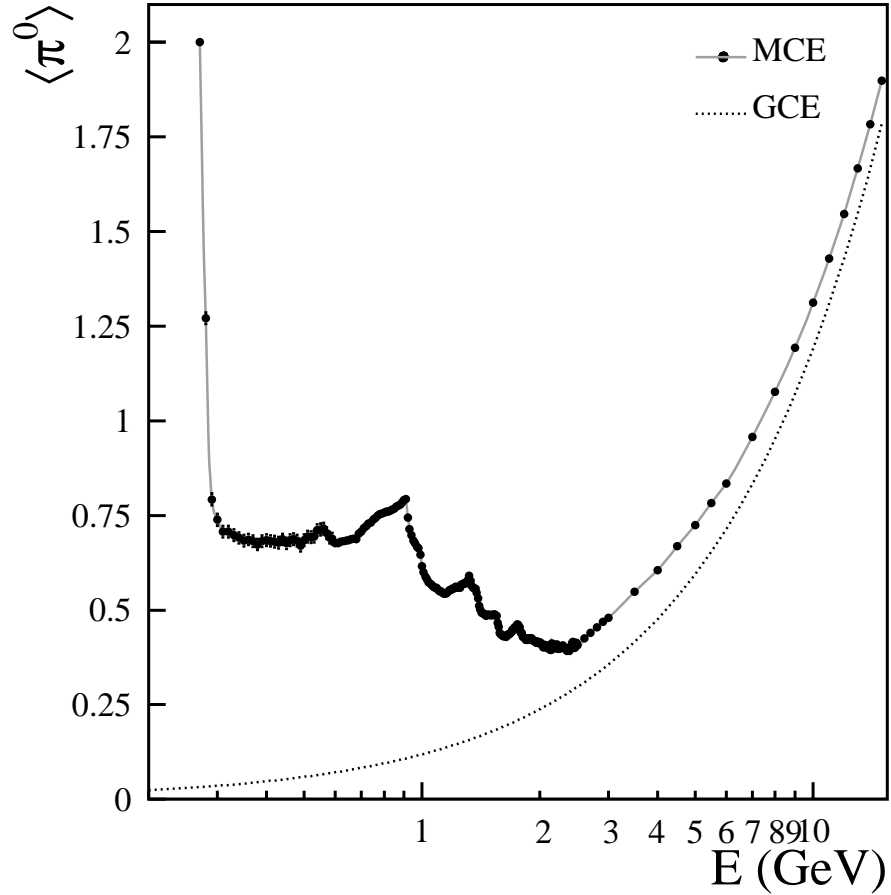}
\caption{
The dependence of the mean multiplicity of $\pi^0$ mesons
on the system energy obtained within the hadron-resonance gas model 
by use of 
the micro-canonical (solid line) and grand canonical (dotted line)
ensembles.
The calculations are performed for T~=~160~MeV and 
$ E \equiv \langle E\rangle_{GCE}$ in GCE and for
 $ E \equiv \varepsilon(T) V$ in MCE.
Connecting lines between the MC-dots are drawn to guide the eyes.
}
\label{fig12}
\end{figure}

\begin{figure}[h!]
\includegraphics[width=0.7\textwidth]{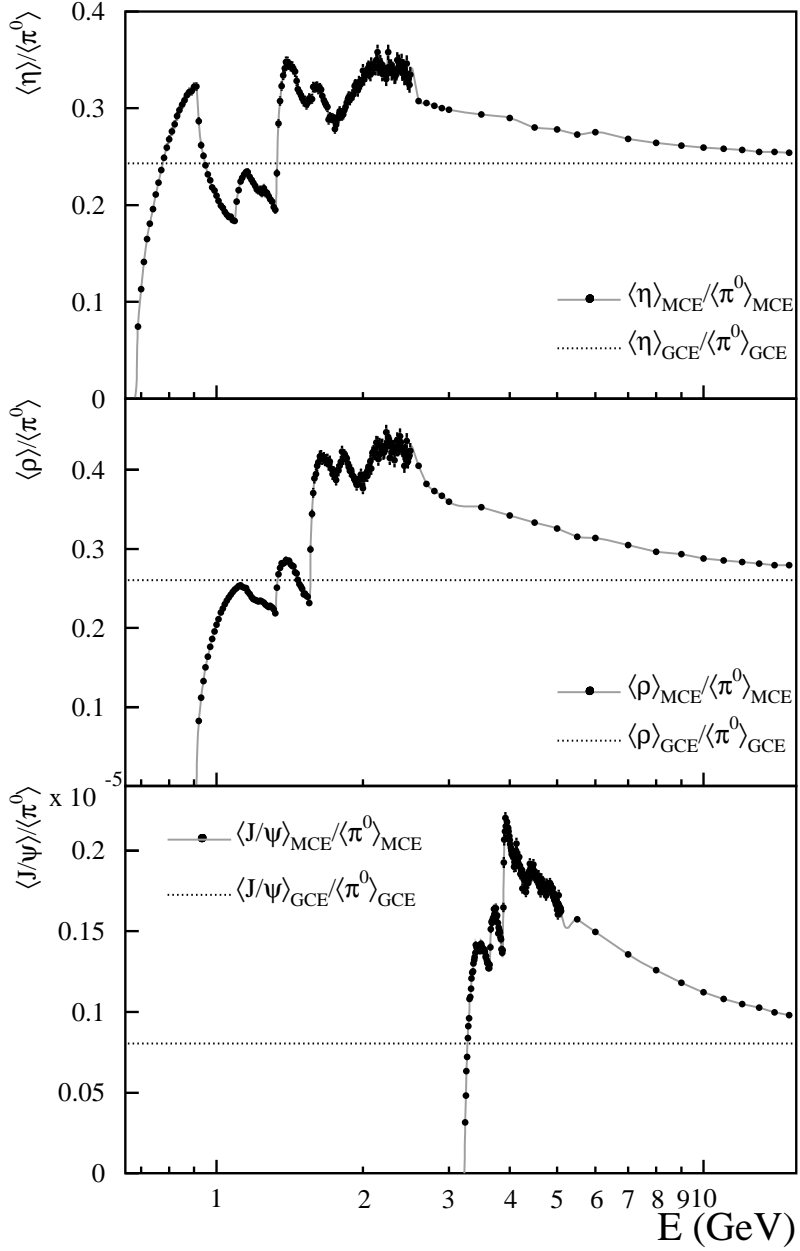}
\caption{
The energy dependence of the ratios 
$\langle \eta \rangle / \langle \pi^0 \rangle$,
$\langle \rho \rangle / \langle \pi^0 \rangle$ and
$\langle J/\psi \rangle / \langle \pi^0 \rangle$
obtained within the hadron-resonance gas model 
by use of 
the micro-canonical (points and solid line) 
and grand canonical (dotted line)
ensembles.
The calculations are performed for T~=~160~MeV and 
$ E \equiv \langle E\rangle_{GCE}$ in GCE and for
 $ E \equiv \varepsilon(T) V$ in MCE.
Connecting lines between the MC-dots are drawn to guide the eyes.
}
\label{fig13}
\end{figure}

In fact, as in the case of the analytical model, we observe a
strong enhancement of the average multiplicities in the MCE with
respect to the GCE near the production threshold. This effect is
larger for lighter particles: about a factor $\sim 10$ in the
ratio MCE/GCE for the second lightest neutral meson $\eta$ and
$\sim 3.5$ for $J/\psi$ meson (see Figs.~\ref{fig7} and
\ref{fig10}). These are remarkable figures, as  the
threshold peaks actually correspond to the ``second'' peaks observed in the
analytical model (see Figs.~\ref{fig2} (left) and (right) as well as
Fig.~\ref{fig3}, (right)). The first peak (the largest one) is
forbidden due to momentum conservation in the MCE that allows no less
than two particles in the final state. A similar effect can be
seen in the analytical model (Fig.~\ref{fig2}) for non-vanishing
charge, because $Q=2$ in the MCE also demands at least two particles
in the final state. One can also see that the peaks in the full
hadron gas have a gradual increase which is typical for the
``second'' peak (compare Figs.~\ref{fig2} (left) and (right)) in the
analytical model.

In full hadron gas
$\pi^+$ and $\pi^-$-mesons play  the role of the massless charged particles
in the
analytical model  considered in Section II.
The ``second peak'' of neutral heavy particle multiplicities is present
similar to that in
Figs.~\ref{fig2} and~\ref{fig3}. This ``second peak'' is
stronger than in an analytical model, because the GCE energy density in
the real hadron gas is a far larger for the same $T=160$~MeV. This is
due to much larger number of
particle species.
An exact momentum conservation makes the whole effect for the
``second peak'' even stronger.

The exact energy-momentum conservation is responsible for a
non-smooth variation of the mean multiplicities as a function of
the energy of the system. In fact, the opening of a new channel at
a total energy $E$ implies a rapid 
change of the behavior of this function. 
In order to
show this, it is appropriate to have a closer look at
the MCE/GCE ratios for $\pi^0,~\eta,~\rho$-mesons. In Fig.~\ref{fig11} a
zoom thereof is shown on the energy range 0.6-1.5 GeV, where
threshold energies for relevant channels are marked. It is evident
that there is a full correspondence between them and the slope
changes. The general pattern can be understood from the analysis
of the analytical massless-heavy particle model. The MCE/GCE ratio
of massless particles gradually decreases to the threshold of heavy
particle production. At the threshold, massless particle yield
drops and heavy ones are enhanced. Likewise, in the full hadron
gas $\pi^0$-mesons play the role of the lightest particles and always
show a drop whenever a new channel is opened. All other particles
behave like ``heavy'' ones near their threshold, but when a new
heavier particle threshold opens up, their MCE/GCE ratio drops,
just as the lightest particles would do. For instance, in the
middle panel of Fig.~\ref{fig11}, the $\eta$ production sets in at the
$\eta \pi^0$ channel threshold and the curve raises until the energy
$m_{\pi^0}+m_{\rho}$ is reached. Here the channel $\rho \pi^0$ is
opened and, just a little later, $\omega \pi^0$ too. The
production onset of these two new heavier particles downgrades
$\eta$ to the role of a light particle and a decrease in the slope
of the ratio $\langle \eta\rangle_{MCE}/\langle \eta
\rangle_{GCE}$ is implied. For little larger energy, the threshold
for the $\eta \eta$ channel entails a new step up in the slope of
the $\eta$ ratio. Then, for $E > m_{\eta}+m_{\rho}$, the channels
$\eta \rho$ and $\eta \omega$ are open and, consequently, both
$\eta$ and $\rho$ ratios increase. Then, for energies larger than
$\sim 1.5$~GeV, the number of allowed channels increases
quickly and the ratio MCE/GCE shows a rocky pattern, which
smoothes out thereafter.

Finally, we turn the quantities which might be, in
general, related
to the experimental results on hadron production in  
collisions of relativistic particles.
Under simplifying assumptions that
the hadron production is statistical and the system
energy is proportional to the collision energy,
hadronic final states can be identified with different 
configurations in the MCE. 
Consequently the dependence of the  ratios of the mean hadron multiplicities
in the MCE on the system energy may be expected to be related
to the collision energy dependence of  these ratios.

The  dependence of  $\langle \pi^0 \rangle$  on the
system energy calculated by use of the MCE and the GCE is 
presented in Fig.~\ref{fig12}. 
There is a smooth and monotonic increase of $\langle \pi^0 \rangle$ in the GCE.
A very different behavior is seen in the MCE. 
For $E <$ 3 GeV the pion yield reveals a non-monotonic, rocky dependence.
In the very nearly of the threshold, $\langle \pi^0 \rangle=2$. This is due to the 
small difference between the mass of $\pi^0$ and $\pi^\pm$ ($\sim 5$~MeV) that prevent charged pions 
from being produced.
For slightly higher energy, as the channel $\pi^++\pi^-$ is open too, $\langle \pi^0 \rangle$ quickly decrease to $\sim 2/3$ as 
said before.

In Fig.~\ref{fig13}, the ratios $\langle \eta \rangle / \langle \pi^0 \rangle$,
$\langle \rho \rangle / \langle \pi^0 \rangle$ and
$\langle J/\psi \rangle / \langle \pi^0 \rangle$ are plotted
as a function of the system energy. 
In the GCE the ratios are independent of the energy, whereas
in the MCE a rocky pattern is predicted.
Only above $E >$~5 GeV the ratios smoothly approach from above the GCE limit. 

In this work we made the assumption of a constant energy-density, that is, in every 
plot the volume is proportional to the energy of the system.
Different hypothesis (for instance an equal constant temperature for GCE and CE and an 
energy-density for MCE: $\varepsilon(T)= E /\langle V \rangle_{CE}$) would lead us to a completely different behavior 
of hadron multiplicities near the threshold with the same thermodynamical limit.
A close comparison with the data could shed some light on which is the right critical quantity 
at hadronization.

%*******************************************************************************
\section{Summary}
%*******************************************************************************
%
We calculated and discussed the threshold behavior of particle
multiplicities in relativistic gases.
The micro-canonical formulations of two models were used.
Firstly, analytical formulas were derived for the gas of massless
and heavy particles.
Second, the results for the hadron-resonance gas are obtained
by use of the Monte-Carlo procedure. 
In both models the ratio of heavy to light particles
near the   threshold is enhanced in comparison to one
in the grand canonical limit.
At low system energies ($E < 3$ GeV) a non-monotonic and rocky
dependence of hadron multiplicities and their ratios on the system energy
is obtained within hadron-resonance gas model. 

The problem whether these unexpected micro-canonical 
threshold effects can be seen in the experimental data
on collisions of relativistic particles is left
for a future study. 
\begin{acknowledgments}
We would like to thank A.P. Kostyuk for useful discussions as well as 
O. Lysak and Marysia Gazdzicka 
for help in the preparation of the manuscript. The work
was partly supported by US Civilian Research and Development
Foundation (CRDF) Cooperative Grants Program, Project Agreement
UKP1-2613-KV-04 and Virtual Institute on Strongly Interacting
Matter (VI-146) of Helmholtz Association, Germany.
\end{acknowledgments}
%
%

%
%
%
%
%\end{document}
\newpage

\end{document}